\documentclass[aps,pra,superscriptaddress]{revtex4}
\usepackage{amsmath}
\usepackage{amssymb}
\usepackage{graphicx}
\usepackage{dcolumn}
\usepackage{bm}

\begin{document}

\title{On the possible wave-packet collapse induced by topological
disconnectivity and experimental suggestions}
\author{Hongwei Xiong \footnote{
xionghongwei@wipm.ac.cn}}
\affiliation{State Key Laboratory of Magnetic Resonance and Atomic and Molecular Physics,
Wuhan Institute of Physics and Mathematics, Chinese Academy of Sciences,
Wuhan 430071, P. R. China}
\affiliation{Center for Cold Atom Physics, Chinese Academy of Sciences, Wuhan 430071,
China }
\date{\today }

\begin{abstract}
The purpose of this paper is to explore the possibility of a
wave-packet collapse induced by topological disconnectivity, based
on the discussions of the wave-corpuscle duality. Several
experimental suggestions are proposed to test this sort of
wave-packet collapse based on quantum interference.


\end{abstract}

\maketitle

\section{\textbf{Introduction}}

Driven by the remarkable quantum manipulations such as the ability
to control individual atoms and the realization of gaseous
Bose-Einstein condensates, the second quantum revolution is under
way. We believe that in the second quantum revolution, a lot of
important techniques will have practical applications, such as
quantum cryptography, and quantum-controlled chemistry, etc. The
advances in quantum manipulation also give us a golden opportunity
to test further the basic concepts of quantum mechanics, whose
meaning has long been debated. A well-known example is the
experimental test \cite{exp-bell} of Bell's inequality \cite{Bell}.
In the last few years, there have been significant advances for the
loophole-free test of Bell's inequality
\cite{Bell-adv,Bell1,Bell2,Bell3}. Recently, an almost ideal
realization of Wheeler's delayed-choice experiment was reported with
a single-photon pulse \cite{Delay}. Most recently, a previously
untested correlation between two entangled photons was measured in
\cite{Gro}, to test an
inequality proposed by Leggett based on a non-local realistic theories \cite%
{leggett-non}. These theoretical and experimental advances have
deepened our understanding of quantum mechanics. In the near future,
we expect that experiments will provide more convincing tests about
the unique concepts in quantum mechanics, such as nonlocality,
wave-packet collapse, and wave-corpuscle duality, etc
\cite{leggett}.

In this work, we consider the possibility of a wave-packet collapse
induced by topological disconnectivity, based on the well-known
wave-corpuscle duality and a simple gedanken experiment.
Several experimental suggestions are proposed to test this
wave-packet collapse based on quantum interference. Although an
experimental investigation of this wave-packet collapse may be quite
challenging, we believe that the rapid advances of quantum
manipulation will make this sort of experiment become promising.

The paper is organized as follows. In Section 2, a possible
wave-packet collapse induced by topological disconnectivity is
discussed based on the wave-corpuscle duality and a gedanken
experiment. In Section 3, the strong disconnectivity and weak
disconnectivity are discussed. In Sections 4 and 5, several
experimental suggestions are proposed to test this wave-packet
collapse for the situations of strong disconnectivity and weak
disconnectivity, respectively. In the last section, a brief summary
and discussion is given.

\section{\textbf{A gedanken experiment and wave-packet collapse induced by
topological disconnectivity}}

We first consider a gedanken experiment shown in Fig. 1. More
realistic experimental proposals will be discussed in Sections 4 and
5. In Fig. 1(a), there are two connected boxes and a shutter. In
each box, there is a trapping potential $V_{1}$ and $V_{2}$,
respectively. The purpose of these two trapping potentials is to
make the wave packet of a particle be highly spatially localized, so
that the direct contact between the particle and the box can be
omitted. It is also required that the direct contact between the
particle and the shutter can be omitted. In an experiment, this can
be prepared and tested by firstly preparing a particle trapped by
the potential $V_{1}$ in the left box. The wave function of the
particle is assumed as $\Psi _{1}$. Then, one closes the shutter to
see whether the closing of the shutter has an influence on the wave
packet of the particle. If the closing of the shutter has negligible
influence on the particle, we think that there is no direct contact
between the particle and the shutter, although the closing of the
shutter has made the left and right boxes become disconnected. It is
similar for a particle in the right box, whose wave function trapped
in the potential $V_{2}$ is assumed as $\Psi _{2}$.

We consider a single particle whose quantum state is a coherent
superposition state, which is given by $\left\vert \Psi
\right\rangle =\alpha \left\vert \Psi _{1}\right\rangle +\beta
\left\vert \Psi _{2}\right\rangle $. Let us discuss the following
problem:

\textit{After two boxes become disconnected by closing the shutter (in the
closing process, there is no external perturbation on the particle and the
direct contact between the shutter and the wave packet of the particle can
be omitted), is there an essential change in the quantum state of the
particle?}

\begin{figure}[tbp]
\includegraphics[width=0.6\linewidth,angle=270]{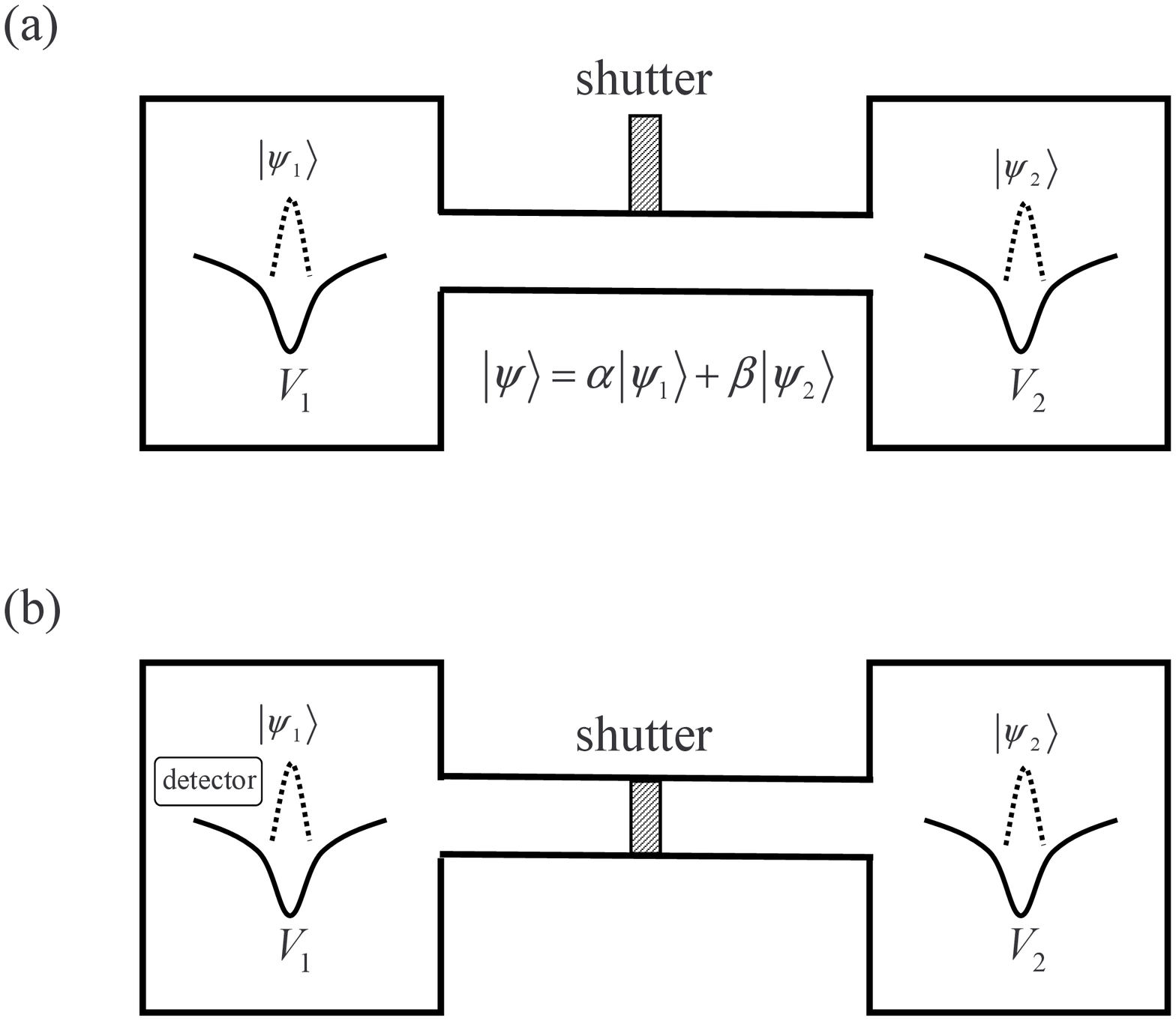}
\caption{Schematic of a gedanken experiment. In Fig. 1(a), the
initial single-particle quantum state is a coherent superposition
state of two spatially-separated wave packets in two connected
boxes. In Fig. 1(b), after the closing of the shutter, two boxes
become disconnected and closed. We consider in the present work the
question whether the closing of the shutter has an essential
influence on the quantum state $\left\vert \Psi \right\rangle
=\protect\alpha \left\vert \Psi _{1}\right\rangle +\protect\beta
\left\vert \Psi _{2}\right\rangle $.}
\end{figure}

For this seemingly simple problem, one is likely to have completely
different answers.

\textbf{Point of view I ------ }\textit{There is no influence on the
single-particle quantum state based on the wave aspect of the particle.}

In our gedanken experiment, it is required that for a particle
(whose quantum state is $\Psi _{1}$) in the left box, the closing of
the shutter has negligible influence on the particle. It is also
required that the closing of the shutter has negligible influence on
the particle (whose quantum state is $\Psi _{2}$) in the right box.
Based on the Schr\H{o}dinger equation, for a particle in the
coherent superposition state $\alpha
\left\vert \Psi _{1}\right\rangle +\beta \left\vert \Psi _{2}\right\rangle $%
, it seems that the closing of the shutter will also have no
influence on the particle. This leads to a predication that there is
no influence on the quantum state when one closes the shutter. This
predication is similar to the situation of a classical wave. One
should note that in these discussions, however, we only address the
wave aspect of a particle. In a sense, in these discussions, there
is no essential difference between a classical wave and a quantum
wave. In the following discussions, we will also consider the role
of the particle aspect of the single-particle quantum state.

\textbf{Point of view II ------} \textit{There is an essential influence on
the single-particle quantum state based on the consideration of the
wave-corpuscle duality.}

The wave-corpuscle duality tells us that the Schr\H{o}dinger
equation describes the wave aspect of the quantum state, while there
exists a statistical bond between the wave aspect and corpuscular
aspect of a particle. Based on the corpuscular aspect, we try to
give here an argument that, there is no single-particle quantum
state which could be a coherent superposition of two wave packets
existing respectively in two topologically disconnected boxes.

We first assume that there is a quantum state $\left\vert \Psi
\right\rangle =\alpha \left\vert \Psi _{1}\right\rangle +\beta
\left\vert \Psi _{2}\right\rangle $ for two topologically
disconnected boxes. In the left box, there is a detector, as shown
in Fig. 1(b). After the detector was switched on, if the detector
records a particle, what the detector measures is a whole particle
because of the corpuscular aspect. This means that the wave packet
$\beta \left\vert \Psi _{2}\right\rangle $ in the right box would
disappear, and this wave packet would appear in the left box. It is
analogous that, if the detector does not record a particle, the wave
packet $\alpha \left\vert \Psi _{1}\right\rangle $ would disappear
in the left box, and appear in the right box. The topological
disconnectivity between two boxes means that the wave packet could
not disappear in one box and appear in another box, if we assume
that there is no extra hidden spatial dimension, or assume that the
wave-packet collapse is a process in the ordinary three-dimensional
space. These discussions lead to a paradox between the wave-packet
collapse (due to the measurement and corpuscular aspect), and the
inhibition of the wave-packet collapse due to the topological
disconnectivity between two boxes. A solution to this paradox is the
suggestion that our initial assumption is invalid, i.e., it is
tempting to assume that there does not exist a quantum state which
is a coherent superposition of two wave packets trapped respectively
in two topologically disconnected boxes.

In the above discussions, we have assumed an inhibition of the
nonlocal wave-packet collapse between two topologically disconnected
boxes. One should note a subtle meaning of the nonlocality of the
wave-packet collapse. Nonlocality has become a key element of
quantum mechanics. For a connected region, the wave-packet collapse
can happen nonlocally and unblockedly. For two topologically
disconnected boxes that a particle in one box can not be transferred
to another box without the breaking of two boxes, however, it is
quite possible that even nonlocality can not transfer a wave packet
from one box to another box. Up to my best knowledge, almost all
theoretical and experimental studies on the nonlocality of quantum
mechanics are carried out for a fully connected region. In
particular, on the experimental side, it seems that the role of
disconnectivity on the nonlocality of quantum mechanics has never
been studied carefully. At least, the previous experiments do not
exclude the possibility of the inhibition of the wave-packet
collapse between two topologically disconnected boxes.

If we agree with the point of view that a particle's quantum state
only exists in one of the two topologically disconnected boxes, we
inevitably get a result that the closing of the shutter would have
an essential influence on the quantum state $\left\vert \Psi
\right\rangle =\alpha \left\vert \Psi _{1}\right\rangle +\beta
\left\vert \Psi _{2}\right\rangle $. These analyses suggest the
following predication:
\begin{eqnarray}
\rho _{i} &=&\left\vert \Psi \right\rangle \left\langle \Psi \right\vert
\notag \\
&\implies &\rho _{f}=\left\vert \alpha \right\vert ^{2}\left\vert \Psi
_{1}\right\rangle \left\langle \Psi _{1}\right\vert +\left\vert \beta
\right\vert ^{2}\left\vert \Psi _{2}\right\rangle \left\langle \Psi
_{2}\right\vert .  \label{collapse}
\end{eqnarray}%
Here $\rho _{i}$ and $\rho _{f}$ are respectively the density matrix before
and after the closing of the shutter.

It is obvious that the probability of finding a particle in the left
or right boxes is the same for the density matrices $\rho _{i}$ and
$\rho _{f}$.
However, there is an essential difference between $\rho _{i}$ and $\rho _{f}$%
, by noting that in the density matrix $\rho _{f}$, there is classical
correlation between two boxes, rather than quantum correlation. In the
density matrix $\rho _{f}$, there is no quantum coherence between $%
\left\vert \Psi _{1}\right\rangle $ and $\left\vert \Psi _{2}\right\rangle $%
. For a particle described by the density matrix $\rho _{f}$, the particle
only exists in one of the boxes with probability $\left\vert \alpha
\right\vert ^{2}$ and $\left\vert \beta \right\vert ^{2}$. If all the
confining conditions (the boxes, the shutter and the trapping potential $%
V_{1}$ and $V_{2}$) are removed, there would be interference between $%
\left\vert \Psi _{1}\right\rangle $ and $\left\vert \Psi _{2}\right\rangle $
after sufficient expansion time for $\rho _{i}$, while there is no
interference for $\rho _{f}$.

Generally speaking, the measurement of a quantum state consists of
two elementary processes: (i) the wave-packet collapse (or
decoherence) process; (ii) a transformation of the collapsed quantum
state to a classical signal we can apperceive. For the situation
discussed in the present work, the closing of the shutter leads to
two topologically disconnected boxes. In a sense, the closing of the
shutter can be regarded as a measurement by two detectors whose
resolution is the region of the left and right box, respectively.
Thus, the closing of the shutter leads to the wave-packet collapse
of a particle, and complete the first process of a measurement. The
second process of the measurement can be completed by switching on
the detectors in two boxes, so that the interaction between the
particle and the detectors gives classical signal we can apperceive.
We want to stress here that even before the detectors in two boxes
are switched on, it is possible that there is already a wave-packet
collapse by the closing of the shutter. In a sense, the closing of
the shutter plays a role of measurement, although further
information is needed so that we know which box the particle exists.
This is a little like the detection of a particle by a detection
screen. When the particle hits the detection screen, the wave-packet
collapse is completed. However, we still need to have a look at the
detection screen to know the location of the particle. Of course, if
the closing of the shutter is regarded as a measurement process,
because there is no spatial contact between the shutter and wave
packets of the particle, it is different from the ordinary
measurement process. In a sense, it is a sort of quantum measurement
process induced by topological disconnectivity.

The different results in the \textit{point of view I and point of
view II} lies in that in the \textit{point of view I} only the wave
aspect of a quantum particle is considered. It is well known that
the wave-corpuscle duality is an essential element in quantum
mechanical principle, and wave-corpuscle duality is the root of the
mystery of quantum mechanics. The Schr\H{o}dinger equation is in a
sense invalid when the corpuscular aspect is considered, such as the
nonlocal wave-packet collapse. Therefore, in a sense, the
\textit{point of view II} is obtained based on more
\textquotedblleft standard\textquotedblright\ quantum mechanics,
compared to that of the \textit{point of view I}.

\section{The definition of strong disconnectivity and weak disconnectivity}

In the \textit{point of view II} about the wave-packet collapse
process given by Eq. (\ref{collapse}), it relies on two assumptions:
wave-corpuscle duality and wave-packet collapse being a process in
the ordinary three-dimensional space. The \textit{point of view II}
is a natural result from the assumption that a particle can not
exist simultaneously in different disconnected regions. We see that
the definition of topologically disconnected regions is necessary to
consider further the problem in the present work.

There are two different topological disconnectivities:

(i) For a particle existing in a region $\Sigma _{1}$, under any
coherent quantum manipulation, if the wave packet of the particle
can not be transformed into another region $\Sigma _{2}$, these two
regions $\Sigma _{1}$ and $\Sigma _{2}$ are regarded as
\textit{strong disconnectivity}.

(ii) For a particle existing in a region $\Sigma _{1}$, there is a
situation that, the manipulation of the spatial wave function alone
can not transform the particle into another region $\Sigma _{1}$.
Together with a manipulation of the internal state, however, one may
transform coherently the wave packet into the region $\Sigma _{2}$.
We call this situation as \textit{weak disconnectivity}.

To make the definition of the weak disconnectivity more clearly, we
give here an example of weak disconnectivity. In Fig. 2, there are
two coherently separated photon wave packets with vertical
polarization and horizontal polarization. In the connection region,
there are horizontal polarizer and vertical polarizer. The cavities
$\Sigma _{1}$ and $\Sigma _{2}$ are disconnected in a sense, because
the left (right) photon with vertical (horizontal) polarization can
not transform into the right (left) cavity without the changes of
the polarization direction. However, if the polarization of the
photon is changed with some quantum manipulations in the cavities,
the wave packet can transform between two cavities. In this
situation, these two cavities are weakly disconnected.

\begin{figure}[tbp]
\includegraphics[width=0.6\linewidth,angle=270]{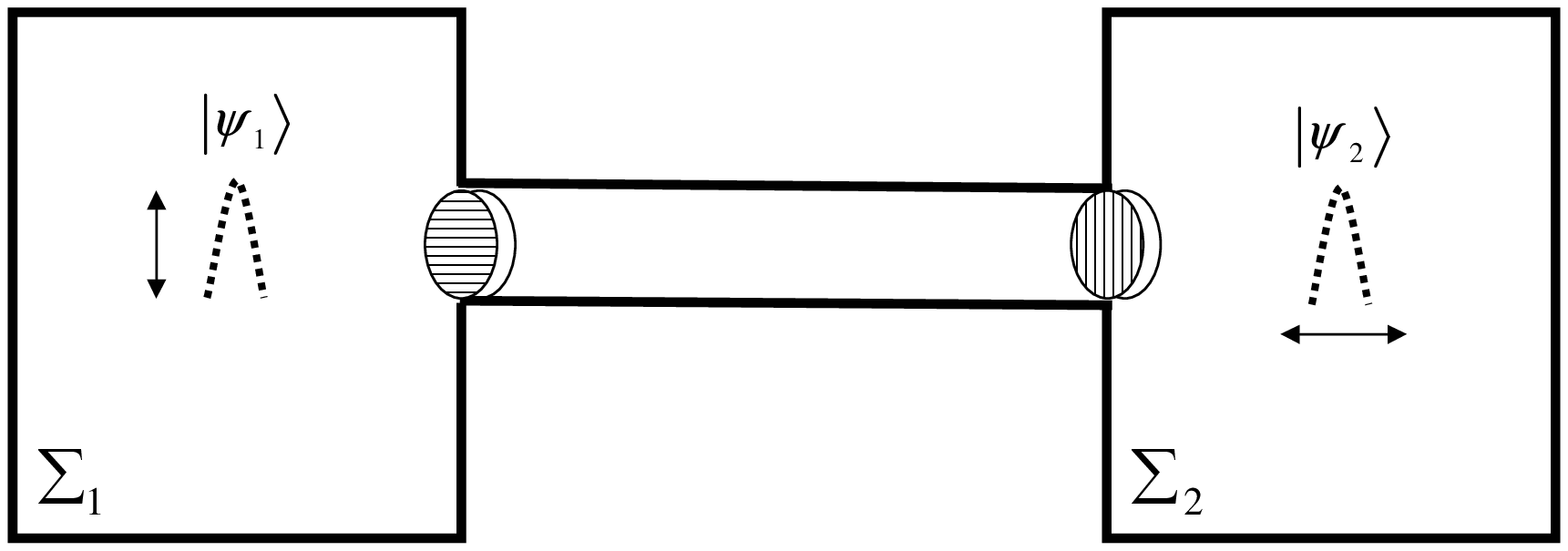}
\caption{Schematic is an example of weak disconnectivity.}
\end{figure}

In my opinion, if the closing of a shutter leads to a strong
disconnectivity, it is possible that there would be a wave-packet
collapse for the coherently superposed quantum state. As for the
situation of weak disconnectivity, at least in my opinion, it is
uncertain what will happen in an experiment. Frankly speaking, I
believe only future experiments could tell us what will happen for
both cases. In the following sections, I will try to give several
experimental schemes to answer these problems in future experiments.


\section{The experimental suggestions for strong disconnectivity}

In this section, we will give several experimental schemes for the
situation of the strong disconnectivity.

\subsection{The experimental suggestion I}

We consider a double-slit interference experiment shown in Fig. 3.
The Young double-slit experiment has been discussed widely for the
complementarity between the corpuscular aspect and wave aspect.
The single-photon pulse is spatially separated after a photon passes
through the double slit in Fig. 3. After the photon passes through
the double slit, the quantum state of the photon is confined in two
optical fibers. For this sort of experiment, there is clear
interference pattern in the detection screen when two shutters $A$
and $B$ are always open. In a real experiment, the connection
between the double slit and the optical fibers may be realized by
using a beam splitter to play the role of the double slit. For this
double-slit experiment, the density matrix is
\begin{equation}
\rho _{1}=\left\vert \Psi \right\rangle \left\langle \Psi \right\vert ,
\end{equation}%
with $\left\vert \Psi \right\rangle =\left( \left\vert \Psi
_{1}\right\rangle +\left\vert \Psi _{2}\right\rangle \right)
/\sqrt{2}$. For a series of single-photon pulses with overall photon
number $N$, the density
distribution on the detection screen is then%
\begin{eqnarray}
n_{1} &=&\left\langle \mathbf{r}|\rho _{1}|\mathbf{r}\right\rangle  \notag \\
&=&N\left[ \frac{1}{2}\left\vert \left\langle \mathbf{r}|\Psi
_{1}\right\rangle \right\vert ^{2}+\frac{1}{2}\left\vert \left\langle
\mathbf{r}|\Psi _{2}\right\rangle \right\vert ^{2}+\mathrm{Re}\left(
\left\langle \mathbf{r}|\Psi _{1}\right\rangle \left\langle \Psi _{2}|%
\mathbf{r}\right\rangle \right) \right] .  \label{density1}
\end{eqnarray}%
The last term is the well-known interference term.

\begin{figure}[tbp]
\includegraphics[width=0.6\linewidth,angle=270]{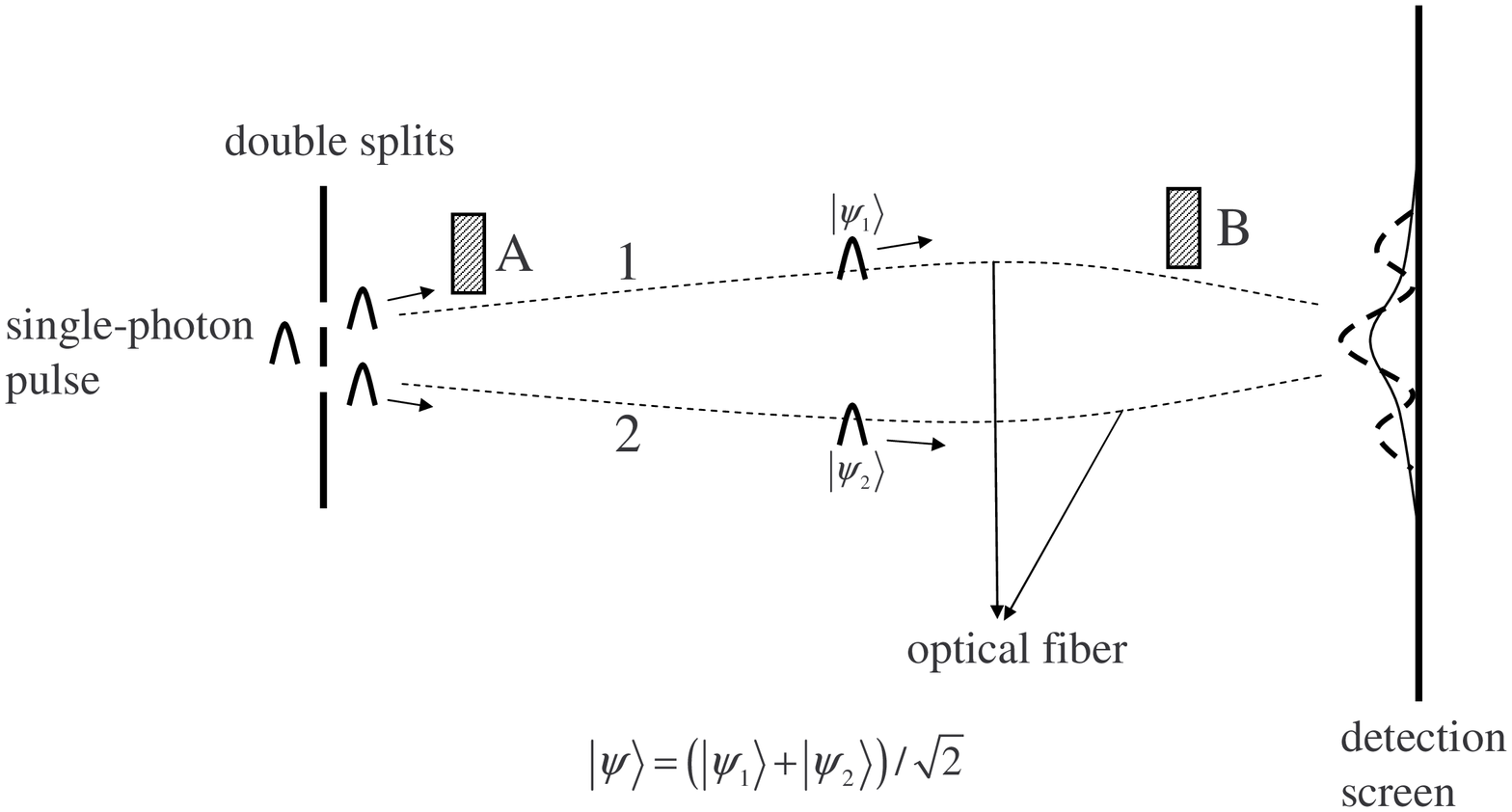}
\caption{Schematic of the Young double-slit experiment. If two
shutters $A$ and $B$ placed along the optical fiber $1$ are always
open, there should be clear interference pattern on the detection
screen shown by the dashed line. When two shutters are manipulated
appropriately, \textit{the point of view I} still predicts the
interference pattern shown by the dashed line, while the
\textit{point of view II} gives the predication of no interference
pattern shown by the solid line.}
\end{figure}

As a comparison, we consider another double-slit experiment by
considering the role of two mechanical shutters $A$ and $B$ placed
along the optical fiber $1$. The mechanical shutters are used so
that one considers experimentally the situation of the strong
disconnectivity. During the propagation of the photon wave packet
between two shutters in the optical fiber $1$, two shutters close
and then open simultaneously. After the two shutters are closed, the
existence region of the quantum state $\left\vert \Psi
_{1}\right\rangle $ and the existence region of the quantum state
$\left\vert \Psi _{2}\right\rangle $ become topologically
disconnected. If the \textit{point of view II} (the wave-packet
collapse given by (\ref{collapse})) is correct,
the closing of the shutter makes the density matrix become%
\begin{equation}
\rho _{2}=\frac{1}{2}\left\vert \Psi _{1}\right\rangle \left\langle \Psi
_{1}\right\vert +\frac{1}{2}\left\vert \Psi _{2}\right\rangle \left\langle
\Psi _{2}\right\vert .
\end{equation}%
Before the possible single-photon pulse arrives at the shutter $B$,
two shutters open simultaneously. For every single-photon pulse,
there is a manipulation of the closing and opening of two shutters
during the propagation of the photon wave packet between two
shutters. In
this situation, for a series of photon pulses with overall photon number $N$%
, the density distribution on the detection screen is%
\begin{equation}
n_{2}=\left\langle \mathbf{r}|\rho _{2}|\mathbf{r}\right\rangle =N\left[
\frac{1}{2}\left\vert \left\langle \mathbf{r}|\Psi _{1}\right\rangle
\right\vert ^{2}+\frac{1}{2}\left\vert \left\langle \mathbf{r}|\Psi
_{2}\right\rangle \right\vert ^{2}\right] .
\end{equation}%
We see that there is no interference pattern based on the \textit{point of
view II}. If the \textit{point of view I} is correct, we expect that there
is still clear interference pattern.

\subsection{experimental suggestion II}

We now consider another experimental suggestion. In Fig. 4, a
single-photon pulse is split into two wave packets by the beam
splitter denoted by BS1. Two spatially separated photon pulses
propagate in the optical fibers along different paths denoted by $x$
and $y$. Two mirrors (denoted by $M$) bring two wave packets
together at a beam splitter denoted by BS2. The light path is
devised so that these two photon pulses arrive at BS2
simultaneously. The phase shifter controls the relative phase $\phi
$ when two photon pulses arrive at BS2. The role of BS2 is shown by
the inset in this figure.
\begin{figure}[tbp]
\includegraphics[width=0.6\linewidth,angle=270]{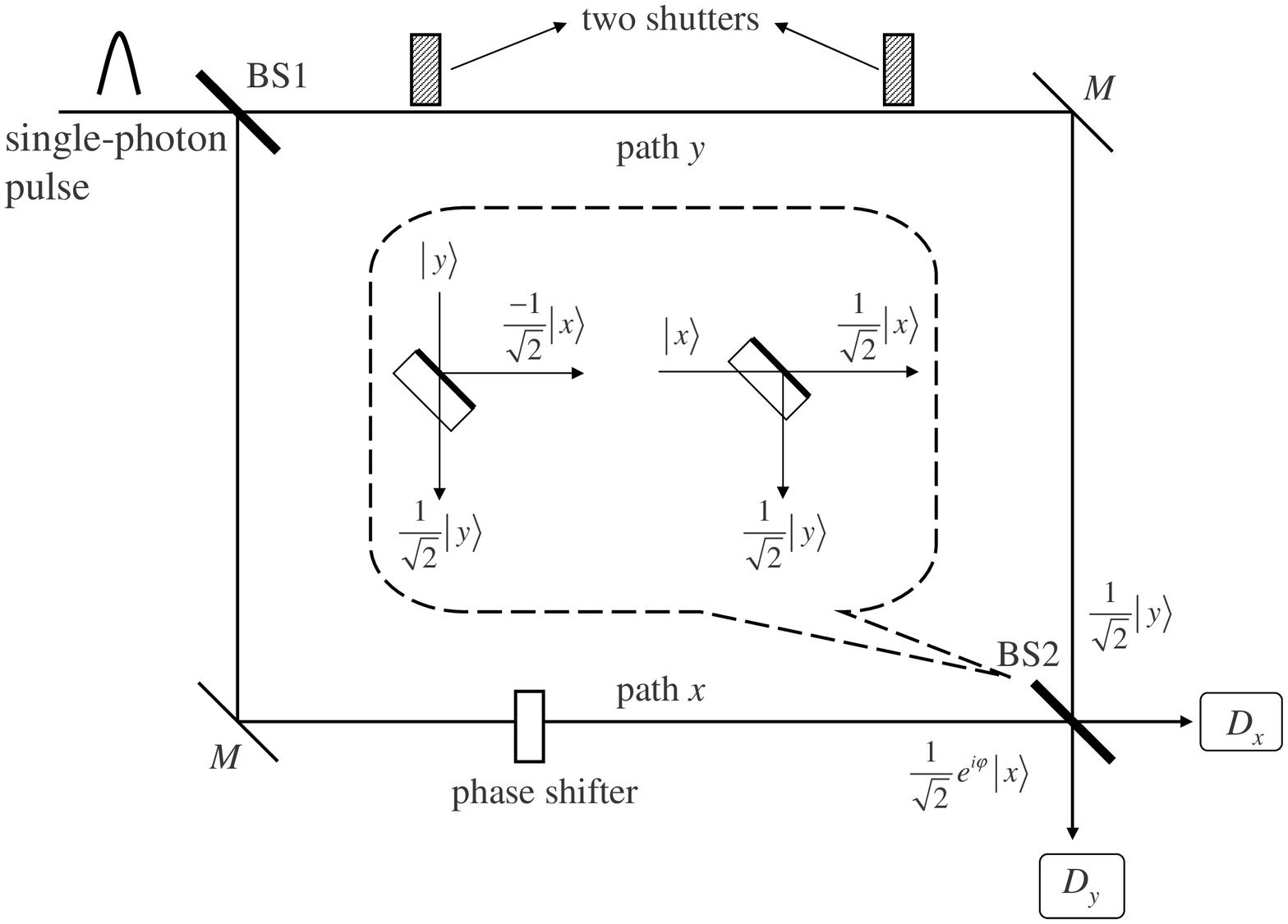}
\caption{Schematic of an experimental suggestion for the situation
of the strong disconnectivity.}
\end{figure}

Without the shutters, after very simple derivation, the density matrix is%
\begin{equation}
\rho _{a}=\left\vert \Psi \right\rangle \left\langle \Psi \right\vert
\end{equation}%
with $\left\vert \Psi \right\rangle =\frac{1}{2}\left( e^{i\varphi
}-1\right) \left\vert x\right\rangle +\frac{1}{2}\left( e^{i\varphi
}+1\right) \left\vert y\right\rangle $. For a series of photon pulses with
overall photon number $N$, the average photon number recorded by the
detectors $D_{x}$ and $D_{y}$ is respectively given by%
\begin{equation}
N_{x}=N\left\langle x|\rho _{a}|x\right\rangle =\frac{N}{4}\left\vert
e^{i\varphi }-1\right\vert ^{2},
\end{equation}%
and%
\begin{equation}
N_{y}=N\left\langle y|\rho _{a}|y\right\rangle =\frac{N}{4}\left\vert
e^{i\varphi }+1\right\vert ^{2}.
\end{equation}%
Because of the interference between the photon's two wave packets,
$N_{x}$ and $N_{y}$ display oscillating behavior with the relative
phase $\varphi $.

Similar to the preceding experimental suggestion, during the propagation of
the pulsed photon between two shutters, two mechanical shutters shown in
Fig. 4 close and then open simultaneously. Based on the \textit{point of view II}%
, the density matrix is%
\begin{equation}
\rho _{b}=\frac{1}{2}\left\vert \Psi _{x}\right\rangle \left\langle \Psi
_{x}\right\vert +\frac{1}{2}\left\vert \Psi _{y}\right\rangle \left\langle
\Psi _{y}\right\vert
\end{equation}%
with%
\begin{equation}
\left\vert \Psi _{x}\right\rangle =\frac{e^{i\varphi }}{\sqrt{2}}\left(
\left\vert x\right\rangle +\left\vert y\right\rangle \right) ,
\end{equation}%
and%
\begin{equation}
\left\vert \Psi _{y}\right\rangle =\frac{1}{\sqrt{2}}\left( -\left\vert
x\right\rangle +\left\vert y\right\rangle \right) .
\end{equation}%
In this situation, we have%
\begin{equation}
N_{x}=N\left\langle x|\rho _{b}|x\right\rangle =\frac{N}{2},
\end{equation}%
and%
\begin{equation}
N_{y}=N\left\langle y|\rho _{b}|y\right\rangle =\frac{N}{2}.
\end{equation}%
Because the photon pulses propagating along the paths $x$ and $y$
become incoherent due to the manipulation of two shutters, $N_{x}$
and $N_{y}$ do not depend on the relative phase $\varphi $.

We see that there is obvious difference based on the
\textit{point of view I} and \textit{point of view II}. In particular, for $%
\varphi =0$, the \textit{point of view I} predicts that $N_{x}=0$, while the
\textit{point of view II} predicts that $N_{x}=N/2$.

\section{The experimental suggestion for weak disconnectivity}

Now we turn to consider an experimental suggestion about the
situation of the weak disconnectivity. The experimental suggestions
in the preceding section have a shortcoming that the response time
of the mechanical shutter should be very short. If one uses a
single-photon pulse, the response time of the mechanical shutter
should be smaller than $L/c$, with $L$ and $c$ denoting the length
of the optical fiber and the velocity of light. This is quite
challenging for a mechanical shutter. For the situation of the weak
disconnectivity, however, the application of the Pockels cell would
overcome the problem of the response time.

Compared with Fig. 4, the path $y$ in Fig. 5 has special design with
two vertical polarizers and two Pockels cells (PC1 and PC2). The
response time of the Pockels cell can be of the order of
$\mathrm{ns}$. The photon from the single-photon source has vertical
polarization. The wave packet of the $n$th single photon propagating
along the path $y$ arrives in succession
the left polarizer, PC1, PC2, and the right polarizer with times $t_{n0}$, $%
t_{n1}$, $t_{n2}$, and $t_{n3}$. We have $t_{n1}-t_{n0}=L_{1}/c$, $%
t_{n2}-t_{n1}=L_{2}/c$ and $t_{n3}-t_{n2}=L_{3}/c$. The Pockels cell
has the role that once a voltage is applied on it, the polarization
of the incident light will be rotated by $90^{\circ }$. The time
sequence of the voltage for two Pockels cells is shown respectively
by solid and dashed lines in Fig. 5. After the time $t_{n1}$, the
polarization of the wave packet of the photon propagating along path
$y$ becomes horizontal. In this situation, we see that between the
time interval $\Delta t$ (shown by the dashed line), the interior
and exterior regions of the box show a sort of weak disconnectivity.
The merit of this experimental scheme lies in that because of the
rapid response of the Pockels cell, it is more feasible than that of
the strong disconnectivity shown in Fig. 4. By checking whether
there is interference phenomena recorded by the detectors $D_{x}$
and $D_{y}$, we will know whether there is a wave-packet collapse
because of the weak disconnectivity.

\begin{figure}[tbp]
\includegraphics[width=0.6\linewidth,angle=270]{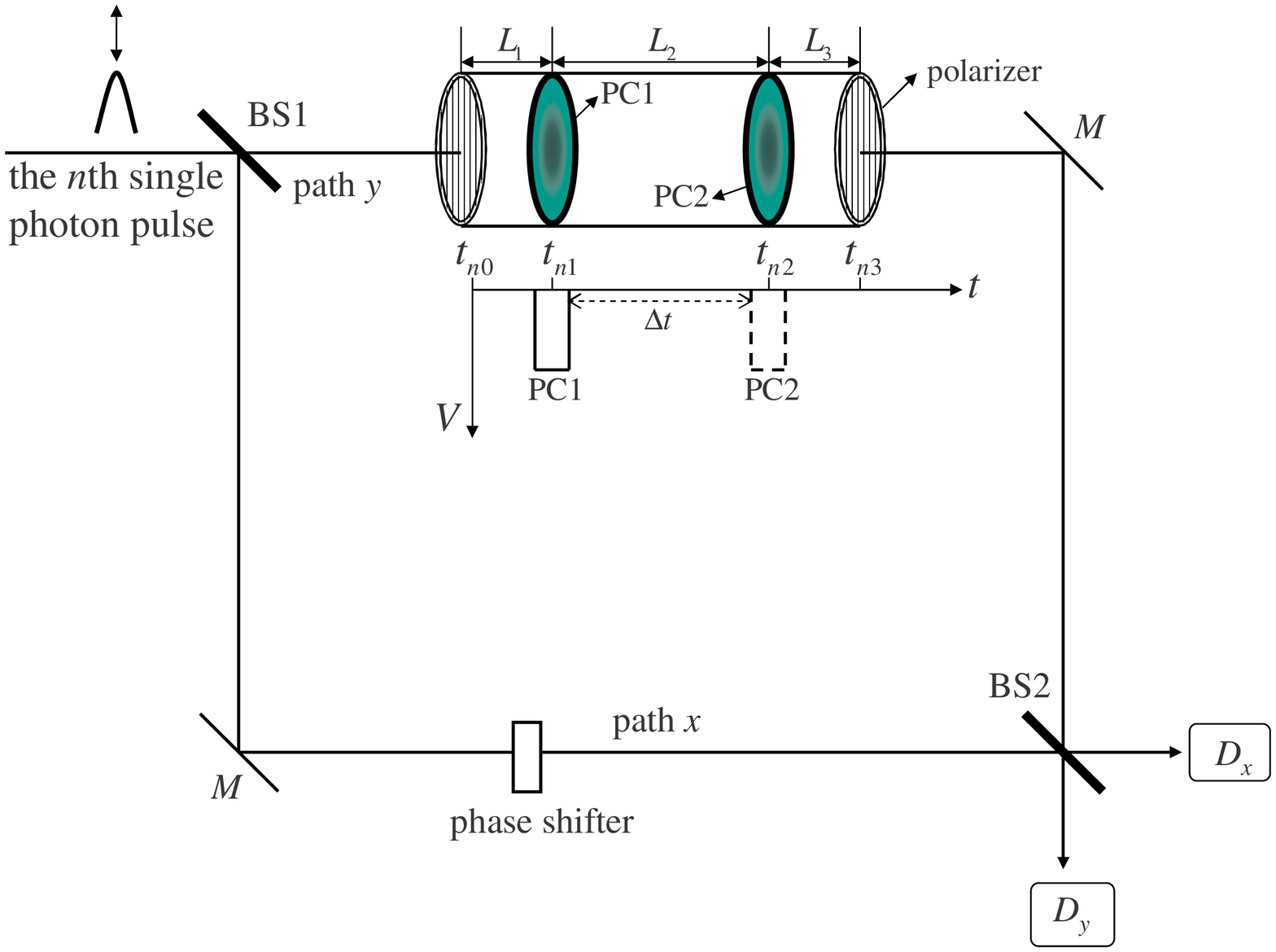}
\caption{Schematic of an experimental suggestion for the situation
of the weak disconnectivity.}
\end{figure}

\section{Summary and discussion}

In summary, we have explored a possible wave-packet collapse induced
by topological disconnectivity. At least for the situation of the
strong disconnectivity, I think that there is a possibility of this
sort of wave-packet collapse, based on the consideration of the
wave-corpuscle duality. Several experimental suggestions are
proposed to test this wave-packet collapse induced by topological
disconnectivity.

One of the main purpose of this paper is to show that there is
interesting physics in future experiments to test the wave-packet
collapse induced by topological disconnectivity. If the
\textit{point of view I} is verified (in particular for the
situation of the strong disconnectivity), the concept of nonlocality
of quantum mechanics can be extended to disconnected regions. This
means that the wave-packet collapse can be regarded as not only
nonlocal, but also in a sense a process beyond the ordinary
three-dimensional physical space. Before the final verification of
the \textit{point of view I}, I could not speak too much about the
meaning of the wave-packet collapse beyond the ordinary
three-dimensional space. If the \textit{point of view II} is
verified, it would establish a new type of wave-packet collapse.
Note that this sort of wave-packet collapse is quite different from
the environment-induced decoherence \cite{decoherence,de1,del2}. In
the environment-induced decoherence, there is a contact between the
wave packet and the environment (or thermal source). For example, in
a beautiful decoherence experiment by using atom interferometry
\cite{atom-dec}, the spontaneously emitted photons play a role of
contact between the spatially separated atomic wave packets and
environment.

Our discussion in this work is different from the delayed-choice
experiments \cite{delay-exp} suggested by Wheeler
\cite{Wheeler,Wh1}. The delayed-choice experiment studied the
interesting quantum interference effect, if the decision for the
observation of a photon in one of the paths is made after the photon
has passed through a beam splitter. In this sort of experiment,
there is also no contact between the wave packet and the apparatus
(a Pockels cell and a Glan polarizing prism in \cite{delay-exp}). In
our gedanken experiment, however, two shutters will produce two
topologically disconnected regions. In fact, after careful searching
the relevant experiments, we find that almost all the quantum
interference experiments do not address the situation of
controllable connected and disconnected regions. The whole region is
always connected in previous Young double-slit experiments, atom
interferometer, etc.

Considering the great challenge of controllable disconnected regions in an
experiment, I think that it is almost impossible to distinguish the \textit{%
point of view I} and \textit{point of view II} by an accidental
experiment. I believe that all previous experiments do not give us
clear evidence to verify which point of view is more reasonable.
Careful division of an experiment is needed to tell us which point
of view is correct. The challenge of this sort of experiment can be
given by a simple analysis. For the scheme shown in Fig. 3 and Fig.
4, if the response time of two electromechanical shutters is a
microsecond, the length of the optical fiber should be larger than
$300$ \textrm{m}. This means that there is severe request on the
quality of the optical fiber and experimental environment, so that
the random phase during the propagation of the photon pulse in the
optical fiber is suppressed largely. One method to overcome this
problem is the development of submicrosecond electromechanical
shutters. Another method is the application of atom interferometer
\cite{inter} to investigate our problem. For this sort of
experiment, two shutters and a cavity should be placed in the vacuum
chamber. After the splitting of an atom, one of the wave packets
passes through the cavity, and the closing of two shutters can make
the cavity become completely closed. For the shutters with response
time of a microsecond and the atoms with velocity of $10$
$\mathrm{m}/\mathrm{s}$, the length of the cavity should be larger
than $10^{-5}$ \textrm{m}. Therefore, it is possible that our
problem could be answered with the development of atom
interferometers.

\begin{acknowledgments}
This work was supported by NSFC under Grant Nos. 10875165, 10804123,
10634060, and NBRPC 2006CB921406. The author thanks very much the
useful discussions with Dr. Hongping Liu, Prof. Baolong Lu and Prof.
Biao Wu \textit{et al}.
\end{acknowledgments}

\end{document}